\documentclass[12pt]{article}

\usepackage{graphicx}
\usepackage{epsfig}
\usepackage[usenames]{color}
\usepackage{epsfig, bm}
\usepackage{latexsym}
\usepackage{bbding}
\usepackage{amssymb}
\usepackage{ifsym}
\usepackage{wasysym}

\RequirePackage{color}
\definecolor{MyDarkGreen}{rgb}{0.02,0.60,0.06}
\newcommand{\red}[1]{{\color{black}{#1}}}

\title{Normalization of peer-evaluation measures of group research quality across academic disciplines}
\author{ 
 {\it R.~Kenna$^{\,1}$} and {\it B.~Berche$^{\,2}$,} \\~\\
$^1$ Applied Mathematics Research Centre,
Coventry University,\\
Coventry, CV1 5FB, England
{}\\~\\
$^2$ Statistical Physics Group,
 Institut Jean Lamour\footnote{Laboratoire associ\'e au CNRS UMR 7198} ,\\
 CNRS -- Nancy Universit\'{e} -- UPVM, B.P. 70239,\\
 F -- 54506 Vand{\oe}uvre l\`es Nancy Cedex, France
{}\\~\\}
\begin{document}
\maketitle

{\Large
  \begin{abstract}
Peer-evaluation based measures of group research quality such as the  UK's Research Assessment Exercise (RAE), 
which do not employ bibliometric analyses, cannot directly avail of such methods to normalize research impact across disciplines.
This is seen as a conspicuous flaw of such exercises and calls have been made to find a remedy.
Here a simple, systematic solution is proposed  based upon a  mathematical model for the relationship between research quality and group quantity.
This model manifests both the Matthew \red{effect} 
and \red{a phenomenon akin to the} Ringelmann effect and 
reveals the existence of two critical masses for each academic discipline:
a lower value, below which groups are vulnerable, and an upper value beyond which the
dependency of quality on quantity reduces and plateaus appear when the critical masses are large.
A possible normalization procedure is then to pitch these plateaus at similar levels.
We examine the consequences of this procedure at RAE for a multitude of academic disciplines,
corresponding to a range of critical masses.

  \end{abstract} }
%
  \thispagestyle{empty}
%
%
  \newpage
%
                  \pagenumbering{arabic}

\section{Introduction}

The assessment of the quality of academic research  has increased in  importance in recent years. 
Evaluation systems such as the UK's {\emph{Research Assessment Exercise (RAE)}\/} form the basis
on which  funding councils and governments decide where to focus investment.
At RAE, academic areas were scrutinised by experts in a multitude of disciplines to determine the proportion of research
which fell into various quality levels. 
On this basis, it is possible to compare between different teams of researchers in a given discipline, 
based at different universities. 

A conspicuous flaw of the RAE is that it does not employ a robust, formal mechanism to normalize results across different academic disciplines.
In a recent interview \red{(Corbyn, 2009)}, Dame Julia Higgins, who has been involved in almost every RAE since
its inception, lamented the  problems of attempting to compare the relative strengths of academic disciplines based on RAE results. 
Higgins pointed out that there is ``no intellectual basis'' for making comparisons
between disciplines and called for a cessation of the myth that some subjects are stronger than certain others.
\red{In the same interview,} it was pointed out  that literal interpretation of the results of 
the most recent RAE would signal
that UK institutions perform better in media studies than in physics, for example.
This can and does have serious implications for the manner in which funding is distributed.

Certainly a straightforward normalization on the basis of averaging over all research teams scrutinised
is not reliable, as different disciplines may have different strengths in a given country. 
While methods exist to normalize bibliometric measures across disciplines, 
these remain controversial (see, e.g.
\red{Alonso {\emph{et al}\/}, 2009;
Leydesdorff and Bornmann, 2010;
Leydesdorff and Opthof, 2010;
Leydesdorff and Shin, 2010;
Opthof and Leydesdorff, 2010;
Tsay, 2009; 
van~Raan {\emph{et al}\/}, 2010;
Waltman {\emph{et al}\/}, 2010;}
 and references therein), 
and are considered inappropriate for usage at 
RAE which focuses on research quality rather than research impact and does not employ citation counts
or bibliometric approaches.
Moreover, studies have claimed that while citation counts may be a reasonable proxy for RAE in some subjects such as biological sciences
and chemistry, they are a weak proxy for a large number of disciplines, including many with good coverage in the Web of Science
(see 
\red{
Harnad, 2008;
Harnad, 2009;
Mahdi {\emph{et al}\/}, 2008;
Oppenheim and Summers, 2008;
van~Raan, 2006a;
}
and references therein). 
\red{Evidence (2009)} have reviewed the appropriateness of bibliometrics in the measurement of research quality
in the UK context.

Therefore, in order to avoid meaningless
comparisons between disciplines in future research evaluation frameworks  and to ensure fairer distributions of resources,
the British funding bodies were called upon to seriously tackle this issue 
in advance of Britain's next  evaluation exercise by finding a way to normalize results across disciplines
\red{(Corbyn, 2009)}.
This paper is an answer to that call.


In Section~2, we outline the workings of the most recent RAE in the UK, called RAE~2008. 
In Section~3, we explain why simple rankings of universities and research groups based on the outcomes of RAE are 
too naive and simplistic to give a true indication of performance, and we introduce  a mathematical model which explains
how research quality is related to the quantity of researchers in  a given RAE team. 
\red{The issue of quantity and quality as cause and effect is also discussed here.}
The model allows for the quantification of the notion of {\emph{critical mass}\/} in research, a notion which has hitherto been
intuitive only. 
In Section~4, our model is used to develop a normalization  method 
which, like the RAE itself, is not reliant on citation counts or bibliometrics.  
Conclusions are drawn in Section~5.

\section{The UK's Research Assessment Exercise (RAE)}

The RAE is a research evaluation exercise which is undertaken in the UK approximately every five years.
Its objective is to determine the quality of research carried out in various teams, in various disciplines, at various 
higher education institutions. The RAE is performed for  four funding councils:
the Higher Education Funding Council for England (HEFCE);
the Scottish  Funding Council (SFC);
the Higher Education Funding Council for Wales (HEFCW);
and the Department for Employment and Learning in Northern Ireland (DEL). 
These four funding councils use the results of the RAE to decide on how to allocate grants for research to 
the institutions which they support. Any higher education institute that receives such funds from any of the four
councils is eligible to participate at RAE. Therefore achieving the right balance of fairness across institutes and 
across disciplines is of paramount importance for the entire UK research community.

For RAE 2008, the research which was evaluated was that which was published or otherwise 
placed in the public domain between 01 January 2001 and 31 July 2007.
In fact, each full-time researcher entered into the process was invited to submit four research outputs
for peer evaluation. Part-time researchers were submitted on a pro rata basis.
The census date was 31 October 2007 and results were announced in December 2008. 
For the purposes of RAE~2008, academia was divided into 67 subject areas called units of assessment (UOAs).
These were grouped into 15 main panels, 
each of which comprised broadly cognate disciplines, whose research specialities have similar approaches.
Work submitted to the evaluation exercise was assessed by panel members and subject-specific experts,
who were drawn from the wider research community.
Cross referencing between sub-panels and panels was used to deal with interdisciplinary research.

At RAE~2008, these experts determined  the proportion of a team's research 
which fell into five quality levels. These were defined as 4* (world-leading),
3* (internationally excellent),  2* (recognised internationally), 
1* (recognised nationally) and unclassified research.
After RAE~2008, HEFCE used a formula based on the emergent quality profiles
to determine the amount of research funding distributed to each university.
That formula associated each
quality rank with a weight so that 4* and 3* research  
respectively received seven and three times the amount of funding 
allocated to 2* research. Research which was ranked as 1* and unclassified research attracted no funding.
This funding formula may therefore be considered as a measure of the quality  of a research team.
Therefore, if we denote  by $p_{n*}$ the percentage of a team's research evaluated as of $n*$ quality, 
we may define the overall quality measure of that team by 
\begin{equation}
s =  p_{4*} + \frac{3}{7}p_{3*} + \frac{1}{7}p_{2*}.
\label{qual}
\end{equation}
In this way, the theoretical maximum possible quality score is $s=100$.

\red{In determining the research-quality proportions $p_{n*}$ for each team, the evaluators considered three aspects: 
research outputs; research environment; and research esteem. Research outputs mostly took the form of publications, but 
could also include patents,  items of software, artefacts, performances, exhibitions or other forms of assessable outputs which demonstrate the quality of research undertaken.
Research environment was exemplified by, for example, 
research income, support funds, research infrastructure, hostings of seminars, workshops, conferences, and of visiting researchers, research students, studentships and research degrees awarded and so on.
Finally, indicators of esteem included awards, prizes, honours, keynote addresses,
and editorial roles. 
It is important to emphasize that an underpinning principle of the RAE is that -- while prominence has been given to research
outputs --  the evaluation is not about  individual researchers, rather concerning 
the whole unit or research group  that is put forward for assessment.
While the relative weighting of outputs, environment and esteem were consistent within a main panel, they were different between panels, complicating inter-panel standardization. }
For example, for Main Panel D (which evaluated submissions in biology, pre-clinical and human biological sciences, 
agriculture, veterinary and food sciences), outputs, environment and esteem were weighted at 75\%, 20\% and 5\%  respectively.
For Main Panel F (pure and applied mathematics, statistics, operational research,
computer science and informatics) these were weighted 70\%, 20\% and 10\%  respectively.
Within a given panel, a small number of submissions were initially assessed, so that sub-panels could develop a common approach and a common
understanding of quality levels. The main panel chairs attended such meetings in an attempt to ensure consistency between UOAs. 
Besides this, no attempt at normalization was implemented, as  no appropriate, bibliometrics-independent method was available
\red{for such a {\emph{team-focused}} exercise}.

Although there is a ``cost weight'' applied to laboratory and clinical subjects,
the formula (\ref{qual}) is the principle determiner of how funding is allocated in England post RAE:
{\red{the quality-related  funding allocated to a group of size $N$ is proportional to 
the product $sN$.}}
But since no formal normalization system was used,  
differences in  stringency across evaluation disciplines can have significant consequences for the way in which funding is allocated 
and this is the source of the concerns expressed \red{by Higgins  (Corbyn, 2009)}. 
Therefore, to ensure equitable allocation of funding  across disciplines a 
 systematic, rigorous, objective normalization approach is essential.

It is important to stress that no form of citation counting was used as a measure of quality at RAE~2008 or any previous exercise
of this type in the UK. 
Moreover, at RAE~2008, only four outputs per researcher are scrutinised, 
rather that their entire portfolios. Furthermore, proposals to employ bibliometrics as direct or proxy measures of quality in Britain's next 
research evaluation framework are controversial, as citations measure impact and not quality
\red{(van Raan 2005)}. 
Therefore it is essential to develop a bibliometrics-independent normalization system which can deal with normalizing 
peer review assessments instead. 
Here this issue is tackled and a method to normalize research evaluation across different academic disciplines is proposed and tested.

\section{The dependence of quality on quantity in research}

To motivate our model, we firstly describe how the usual, naive interpretation of research evaluation results is 
expressed mathematically, and why this interpretation is flawed. We then propose a more sophisticated model which 
remedies these flaws by taking into account interactions between researchers. We report upon rigorous statistical testing of that 
model, upon which our subsequent normalization method and analysis are based.

To begin, we use the example of the physics UOA at RAE~2008 and plot in Fig.1(a) the quality scores
as given by (\ref{qual}) for each institution alphabetically listed. Clearly, and as one would expect, there is a 
scatter about  a mean value. One naively interprets this as meaning that groups whose quality measures
are below the mean are performing relatively weakly (below what might be expected) and those above the mean are   strong.
We next interpret this naive viewpoint mathematically. We then introduce our more sophisticated model which will show that
the naive viewpoint is incorrect.

We denote the research strength of the $i^{\rm{}th}$ member of the $g^{\rm{th}}$ research group\footnote{Note 
that we use the word ``group'' here to denote  the collection of 
researchers at a given university who were submitted to RAE in a given UOA. This is not synonymous with the department whence
they were drawn because not all departmental members may have been submitted to RAE or a submission may draw from researchers interacting
across different departments. The word ``group'' in this sense is also not synonymous with research centre, as such an entity may be 
involved in submissions across more than one UOA.} in a given discipline by ${a_g}_i$.
The naive view is that the total strength $S_g$ of group $g$ is simply given by the sum of the strengths of its individual members, so that
\begin{equation}
 S_g = \sum_{i=1}^{N_g}{} {a_g}_i = N_g {\bar{a}}_g,
\end{equation}
where $N_g$ is the number of individuals in the group and where ${\bar{a}}_g$ is their mean strength.
We now define the {\emph{quality}} $s_g$ of the entire group as the mean strength per head so that 
$s_g = S_g/N_g$. Then we have the {\emph{naive}} (and, as we shall see, {\emph{erroneous}\/}) conclusion that
\begin{equation}
 s_g = {\bar{a}}_g,
\label{wrong}
\end{equation}
or the quality of the group is given by the mean strength of its individual members.
This is the simple mathematics behind Fig.1(a).
Even worse (as we shall see) is the interpretation that the average strength of the individual members of group $g$
may be  approximated by  RAE measured group quality \red{(Oppenheim and Summers 2008)}.
We will show that such an interpretation is dangerously wrong.

Our first hint as to the inappropriateness of model (\ref{wrong}) comes from Fig.1(b), where quality is plotted against the 
quantity $N$ of researchers in each group for physics. Clearly there is a correlation between quality and quantity up to 
a certain group size (about $N=30$ for physics), beyond which quality tends to plateau. This behaviour is missed by the simple model (\ref{wrong}) and an acceptable model must be able to account for it.

\begin{figure}[!t]
\begin{center}
\hspace{-1cm}
\includegraphics[width=0.45\columnwidth, angle=0]{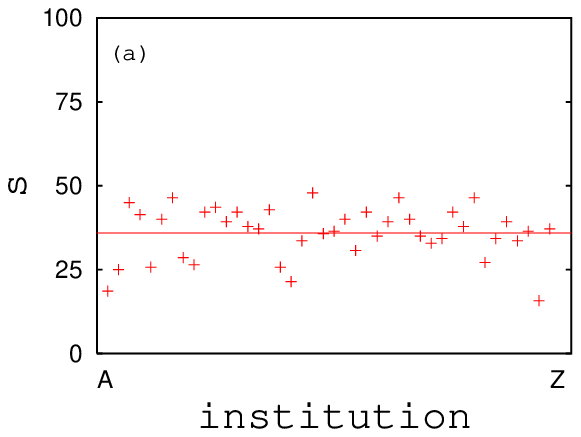}
\includegraphics[width=0.45\columnwidth, angle=0]{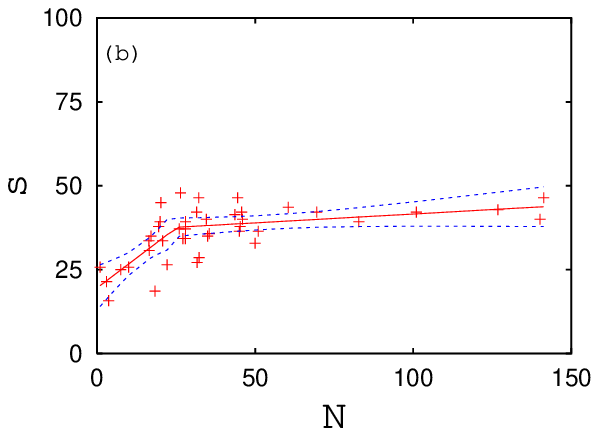}
\caption{Quality measurements for each of the 42 UK physics groups (a) plotted alphabetically
according to university name
and (b) plotted against group size. In panel (a), the solid line is the mean quality of these 42 groups.
In (b), a correlation between quality and groups size is evident. The solid line  is a
piecewise linear regression best-fit to the data and the  dashed curves  represent 95\%  
confidence intervals for this fit.}
\end{center}
\end{figure}

By considering research groups as {\emph{complex systems}\/}, in which the interactions between individuals
play important roles, a mathematical model for the dependency of research quality on group size
was developed by Kenna and Berche \red{(2010; 2011)}. According to the model, the strength of research depends both on the quantity
$N$ and quality of researchers in a research group {\emph{and}} on the number and strengths of interactions between them.
We represent by ${b_g}_{\langle{i,j}\rangle}$ the strength of interaction between the $i^{\rm{th}}$ and $j^{\rm{th}}$
individuals of the $g^{\rm{th}}$ group. 
The potential number of two-way communication links between researchers in this group is $N_g(N_g-1)/2$. 
If the group is sufficiently compact, all of these communication links may be active and Eq.(\ref{wrong}) should be replaced by 
\begin{equation}
 S_g = \sum_{i=1}^{N_g}{{a_g}_i} + \sum_{\langle{i,j}\rangle=1}^{N_g(N_g-1)/2}{{b_g}_{\langle{i,j}\rangle}} = N_g \bar{a}_g 
+ \frac{1}{2} N_g(N_g-1) \bar{b}_g,
\label{correct}
\end{equation}
where  $\bar{b}_g$ is the mean intra-group interaction strength.
However, meaningful two-way communication can only be  carried out between a limited number of researchers, which we denote by $N_c$.
The value of $N_c$ can and, as we shall see, does depend on the academic discipline involved.
Beyond this group size, the group fragments into subgroups,  the mean size of  which {\red{we write as $\alpha_g N_c$.}}
Of course, there is also communication between subgroups and we denote the average strength of these by $\beta_g$. 
The total strength of the $g^{\rm{th}}$ group is then given by the combined individual, intra-group interaction and 
inter-subgroup interaction strengths as
\begin{equation}
 S_g =  N_g \bar{a}_g + \frac{1}{2} N_g(\alpha_g N_c-1) \bar{b}_g + \frac{\beta_g}{2}\frac{N_g}{\alpha_g N_c}
\left({ \frac{N_g}{\alpha_gN_c}-1}\right).
\label{correct2}
\end{equation}

In Eqs.(\ref{correct}) and (\ref{correct2}), the parameters $\bar{a}_g$, $\bar{b}_g$, $\alpha_g$ and $\beta_g$ represent
features of the $g^{\rm{th}}$ group. 
Averaging these values across all groups of that size gives a representation of the
expected behaviour of research groups in the discipline to which group $g$ belongs.
We denote these means as  $a$, $b$, $\alpha$ and $\beta$, respectively, to
 find that the expected strength $S$ of a research group of size $N$ in a given discipline is
\begin{equation}
 S = \left\{ \begin{array}{ll}
             \left({ a - \frac{b}{2}}\right)N + \frac{b}{2} N^2 &  {\mbox{if $N \le N_c$}} \\
             \left({ a + \frac{b}{2}(\alpha N_c-1) - \frac{\beta}{2\alpha N_c}}\right)N + \frac{\beta}{2\alpha^2 N_c^2} N^2 &  {\mbox{if $N \ge N_c$}}.
             \end{array}
     \right.
\label{Ncccc}
\end{equation}

To simplify this expression we introduce the discipline-dependent parameters
\begin{eqnarray}
 a_1 & = & a - \frac{b}{2}, \label{aa} \\
 b_1 & = & \frac{b}{2}, \label{b2}\\
 a_2 & = & a + \frac{b}{2}(\alpha N_c-1) - \frac{\beta}{2\alpha N_c}, \\
 b_2 & = & \frac{\beta}{2\alpha^2 N_c^2} \label{bb}.
\end{eqnarray}
It is interesting to note that the last of these formulae gives an inverse proportionality between $b_2$ and $N_c^2$.
This will mean that for disciplines with large values of $N_c$,
the plots of group quality $s$ versus group size $N$ will have 
small slopes in the $N>N_c$ regime.
Finally, {\red{defining research quality as the strength per head,
\begin{equation}
s=\frac{S}{N},
\label{strengthquality}
\end{equation}
Eqs.}}(\ref{aa})--(\ref{bb}) yield a piecewise linear expression for group quality as a function of quantity:
\begin{equation}
 s = \left\{ \begin{array}{ll}
             a_1 + b_1 N &  {\mbox{if $N \le N_c$}} \\
             a_2 + b_2 N &  {\mbox{if $N \ge N_c$}}.
             \end{array}
     \right.
\label{Nc}
\end{equation}
From this expression, it is clear that for  $N>N_c$, two-way communication between {\emph{all}} team members ceases as the dominant 
driver of research quality. Instead there is a weaker dependency of $s$ on $N$ .
This phenomenon is akin to phase transitions in physics and the model (\ref{Nc}) is reminiscent of mean field theories
of critical phenomena.
We can consider the breakpoint  $N_c$  as a demarcation  between what we term ``small/medium'' and ``large'' groups.
It measures the average number of colleagues with whom a given individual can collaborate in a meaningful sense within a
research team, in a given discipline.

The model (\ref{Nc}) was extensively and rigorously tested using a variety of statistical methods 
\red{(Kenna and Berche, 2010; 2011)}.
In particular, the $P$-values for the null hypothesis based on (\ref{wrong}) that there is no underlying correlation between $s$ and $N$ was
tested for 24 academic disciplines and was rejected at the 5\% level in each case. 
Further hypothesis testing included tests for the existence of the breakpoint and for the reduction of strength of the dependency of
quality and quantity as $N$ increases through $N_c$.
Thus very strong evidence for the validity of model (\ref{Nc}) was presented.
We refer to $N_c$ as the {\emph{upper critical mass}\/} of a given academic discipline. It marks the 
group size beyond which increased staff numbers do not yield significant increase in research quality.

A {\emph{lower critical mass}\/} was also introduced by \red{Kenna and Berche (2010; 2011)}. This is the minimum size a 
research group must achieve for it to remain viable. 
This more closely corresponds to the traditional notion of critical mass in research \red{(Harrison, 2009)}.
Denoting it as $N_k$, a relationship between the 
two critical masses was established  as
\begin{equation}
 N_c = 2N_k.
\end{equation}
This relationship (\ref{b2}) was also borne out in the analyses of \red{Kenna and Berche (2011)}. 
Furthermore, of the subject areas analysed,
those  which had $N_c >14$ tended to have $b_2$ values compatible with zero while
the areas with smaller breakpoints tended to have positive slopes on the right of the curve. 

Having quantified the notion of critical mass in research, and demonstrated that there are in fact two of them, 
we may then introduce a discipline-dependent classification system as follows. We call a research group
\begin{eqnarray*}
 {\mbox{small or subcritical if}} &  &  N \le N_k, \\
 {\mbox{medium if}} &  & N_k \le N \le N_c, \\
 {\mbox{large or supercritical if}} &  & N \ge N_c.
\end{eqnarray*}
The {\emph{upper}\/} critical mass estimates resulting from applying piecewise linear fits to the model (\ref{Nc}) using 
these RAE quality scores are listed in Table~1 for a variety of disciplines.

\begin{table}[t!]
\caption{The {\emph{upper}\/} critical mass estimates from \red{Kenna and Berche (2010; 2011)} are given in the second column.
The third column contains the weighted supercritical mean RAE raw quality scores $\tilde{s}_>$, 
the average of which is $\hat{s}_> =43.2$.
Reweighting each $\tilde{s}_>$ value to this average normalizes the quality scores by the factor 
given in the final column.
}
\begin{center}
\begin{tabular}{|l|r|r|r|} \hline \hline
Subject                                  & $N_c$       &$\tilde{s}_>$&$\hat{s}_>/\tilde{s}_>$   \\
                                         &                &          &               \\
\hline
Computer science \& informatics            & $49.0 \pm 10.0$&  $57.5$  &  0.75                      \\
Economics \& econometrics                & $10.7 \pm 2.7$ &   $51.9$ &   0.83                       \\
English language \& literature          &$31.8 \pm 2.8$  &  $51.7$  &    0.84                      \\
Philosophy \& theology                   &$19.0 \pm 2.9$  &  $48.1$  &    0.90                      \\
Medical sciences                         &$40.8 \pm 8.0$  &  $47.6$  &  0.91                       \\
Chemistry                                & $36.2 \pm 12.7$& $46.7$   &   0.93                   \\
History of art, performing arts,         &$ 8.9 \pm 1.6$  &  $45.3$  &  0.95                       \\
communication studies \& music           &                &          &                       \\
Archaeology                               & $17.0 \pm 2.4$ &  $45.5$  &  0.95                     \\
History                                  &$24.9 \pm 4.5$  &  $45.7$  &    0.95                     \\
Geography, Earth \& environment          & $30.4 \pm 2.8$ &  $44.5$  &  0.97                     \\
Law                                      & $30.9 \pm 3.8$ &  $42.0$  &  1.03                     \\
Applied mathematics                      & $12.5 \pm 1.8$ &  $41.9$  &   1.03                      \\
Architecture \& planning                 & $14.2 \pm 2.8$ &  $41.5$  &  1.04                     \\
Physics                                  & $25.3 \pm 4.7$ &  $40.2$  &  1.07                      \\
Pure mathematics                         & $ \le 4 $      &   $40.4$ &   1.07                       \\
Politics \& international studies        &$25.0 \pm 4.1$&$39.5$  &  1.09                       \\
Education                                &$29.0 \pm 4.4$  &  $38.7$  &    1.12                      \\
Sociology                                &$14.0 \pm 3.1$  &  $38.7$  &  1.12                     \\
Art \& design                            &$25.0 \pm 7.4$  &  $37.9$  &   1.14                      \\
Biology                                  & $20.8 \pm 3.1$ &  $37.1$  &  1.16                      \\
French, German, Dutch  \& Scandinavian   & $ 6.5 \pm 0.8$ &  36.5    &   1.19                       \\
Agriculture, veterinary \& food sciences & $ 9.8 \pm 2.7$ &  $31.8$  &   1.36                     \\
\hline
Average   $\hat{s}_>$                    &               &  $43.2$  &                         \\
\hline
\hline 
\end{tabular}
\end{center}
\end{table}

To summarise, we have developed a mathematical model for the relationship between the quality of research and the quantity of
researchers in a group. This model is successful in describing the results coming from the British RAE~2008
peer evaluation exercise (see Fig.1(b)) 
and is superior to the naive viewpoint that quality is independent of quantity (Fig.1(a)). 
The model allows for the categorisation of research groups as ``small'', ``medium'' and ``large'' and
for each discipline there are two critical masses which separate these categories. 
Research quality increases linearly with group size for small and medium  groups, but this relationship
reduces significantly for large groups. Indeed, the quality of research for such groups in disciplines 
 with relatively large critical masses does not increase significantly with increasing mass.

In going from an interpretation based on Fig.1(a) to one based on Fig.1(b), we have encountered an example
of what is known in the literature as the Matthew effect or cumulative advantage\footnote{The name comes from the Gospel 
according to St. Matthew, which states 
``For unto every one that hath shall be given, and he shall have abundance: but from him that hath not 
shall be taken away even that which he hath''.
}. This is a phenomenon whereby ``the
rich get richer and the poor get poorer'' \red{(Merton, 1968)}. In our case, bigger groups tend to be more successful,
and therefore may tend to grow even bigger.
Our model explains the Matthew effect for group quality as a function of quantity 
in the context of peer research evaluation exercises such as the RAE, at least up to the breakpoint.

The Matthew effect has been observed before in a similar context, namely in the dependency of {\emph{impact}\/} 
(in the form of numbers of citations) on {\emph{volume}\/} of research (in the form of numbers of publications).
Katz \red{(1999;2000)} has shown that this dependency is described by a power-law scaling relationship across research disciplines,
institutes and nations. Such behaviour is characteristic of self-similar systems in physics. 
\red{van Raan (2006b) and Costas {\emph{et al}\/} (2009) }
extended these scaling relationships  to smaller
 entities, namely research groups and individuals, and presented empirical 
results on the statistical properties of standard bibliometrics applied at these levels.
These studies were further extended by van~Raan \red{(2006c)} 
to an analysis of the differences in statistical properties of top- and lower-performance groups 
and to analyse at the levels of universities  \red{(van~Raan, 2008)}.
Each of these studies support the notion of cumulative advantage in the science system, at least insofar 
as the relation between number of citations and number of publications is concerned.

A somewhat opposite, breaking phenomenon, which to our knowledge has not been discussed in the current context, 
is the Ringelmann effect \red{(Ringelmann, 1913)}. 
\red{This is a phenomenon whereby the productivity increases as groups grow in size, but where the gain reduces 
for each new group member added (i.e., the rate of increase in productivity per group member drops as the group size increases).}
The Ringelmann effect was originally thought to be attributable either to decreasing motivation of individuals or to 
\red{coordination} problems as team size increases. 
Following experiments by \red{Ingham {\emph{et al}\/} (1974)}, the Ringelmann effect is nowadays mostly 
considered to be a motivational effect \red{-- the contribution per head decreases as the team size increases. 
Here we observe a similar, yet distinct effect which is clearly organisational rather than motivational; 
as the group size transcends the upper critical mass,} a phase transition occurs when the team size exceeds the number of individuals with whom one can meaningfully communicate. \red{Rather than the decrease in the contribution per head (the quality 
in our case) associated with the Ringelmann effect, our model exhibits a decrease {\emph{in the rate of change}} 
of quality per head,  i.e., a {\emph{reduced increase}} or {\emph{plateau}} in the quality itself. }

\red{A group-growing} mechanism whose causality is the reverse to the main one proposed by us
\red{may also be envisioned (Katz, 2005; van Raan, 2006d)}. 
\red{It has been} argued that successful groups are able to attract more research funding and thus to enlarge further, 
i.e. that quality drives quantity rather than the reverse. 
However,  since there is no ``breaking mechanism'' in \red{this} causal direction, 
if this were the primary driver for the growth of groups, one would expect it would lead to a continued Matthew effect  
and a sustained increase of quality with quantity up to the maximum possible level of $s \approx 100\%$. 
No research team at RAE came close to such a score, with the best teams achieving approximately half this. 
Moreover, the reverse causal mechanism cannot explain the existence of breakpoints or the onset of the 
Ringelmann\red{-type} effect, which our statistical analyses have so clearly established 
\red{(Kenna and Berche, 2010; 2011)}.
Our model (\ref{Nc}) explains both the linear increase of quality with quantity up to the breakpoint 
(the Matthew effect) and the reduction of this phenomenon beyond the breakpoint (\red{akin to} the Ringelmann effect), 
as well as the existence of the breakpoint itself. 
It also allows for the \red{opposite} causal direction, but only as a second-order, sub-dominant, feedback mechanism: 
in our model, increasing quantity driving increasing quality is the dominant growth mechanism and this is supported by extensive statistical analyses \red{(Kenna and Berche 2010; 2011)}.

Having established the appropriateness of the model (\ref{qual}) for the description of the dependency of
research quality on group quantity, we next move on to describe how it may be utilised to normalize
RAE scores across different disciplines.

\section{Normalization across research disciplines}

\begin{figure}[t]
\begin{center}
\includegraphics[width=0.49\columnwidth, angle=0]{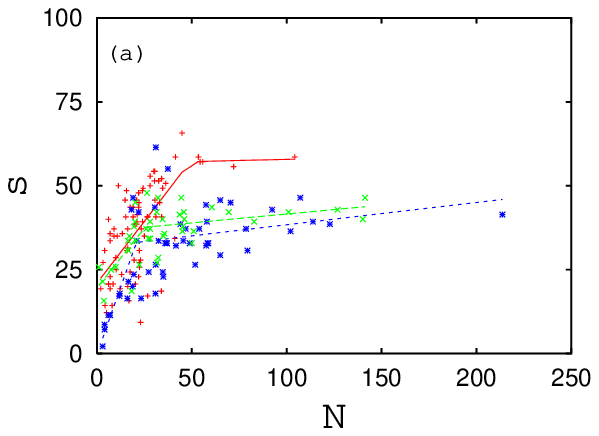}
\includegraphics[width=0.49\columnwidth, angle=0]{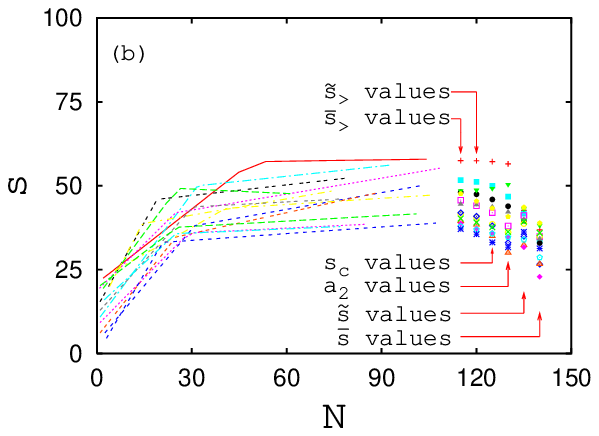}
\caption{(a) Plots of team quality against quantity for
computer science \& informatics~({\color{red}{+}}),
physics~({\color{green}{$\times$}}), 
and biology~({\color{blue}{\textasteriskcentered}}) as measured at RAE.
The three curves are the corresponding piecewise linear fits to model (\ref{Nc}).
(b) The three fits of part (a) together with those for  
English language \& literature~({\color{cyan}{$\blacksquare$}}),
philosophy \& theology~({\color{black}{$\bullet$}}),
history~({\color{magenta}{$\Box$}}),
archaeology~({\color{green}{$\blacktriangledown$}}),
architecture \& planning~({\color{yellow}{O}}),
law~({\color{blue}{$\Diamond$}}),
politics \& international studies~({\color{BurntOrange}{$\triangle$}}),
geography, Earth \& environmental studies~({\color{Gray}{$\blacktriangle$}}),
medical sciences~({\color{yellow}{$\bullet$}}),
education~({\color{magenta}{$\blacklozenge$}}),
art \& design~({\color{cyan}{$\pentagon$}}).
The statistics in the right part of the plot are discussed in the text.
}\end{center}
\end{figure}

\begin{figure}[t]
\begin{center}
\includegraphics[width=0.49\columnwidth, angle=0]{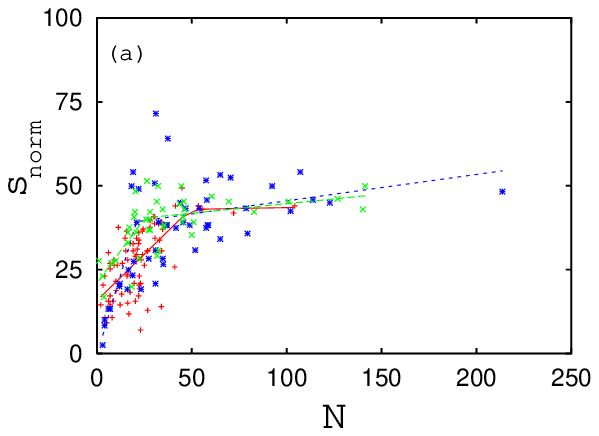}
\includegraphics[width=0.49\columnwidth, angle=0]{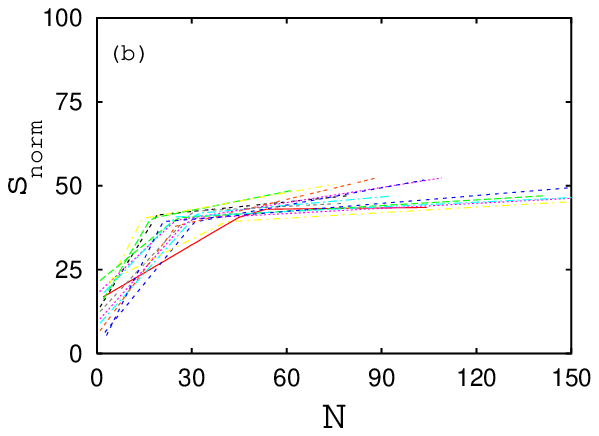}
\caption{(a) The result of the normalization procedure on the three sets of data and corresponding fits of Fig.2(a). (b) The post-normalization fits for the disciplines of Fig.2(b).}
\end{center}
\end{figure}

According to the mathematical model introduced in Section~3 and supported by statistical analyses 
\red{(Kenna and Berche, 2010; 2011)},  
plots of the  dependency of group quality on group quantity are expected to exhibit saturation to the right of the breakpoint
for disciplines with large critical mass values. 
The existence of the breakpoint indicates that groups to the right are performing maximally - 
further increase in group size does not significantly increase group quality. This observation forms the 
crux of our normalization scheme: it is sensible to peg maximally performing groups in different disciplines
at similar levels.

In Fig.2(a) the  research quality scores $s$ are plotted against team sizes $N$ for three different areas, namely  
computer science and informatics, physics, and biology.
Piecewise linear regression fits to the model  (\ref{Nc}) indicate that each of these areas have relatively large critical 
masses and supercritical slopes which are compatible with zero, thus facilitating  comparisons between them\footnote{
In fact three critical masses were identified for computer science and informatics,
indicating that that discipline is actually an amalgam of separate subdisciplines \red{(Kenna and Berche, 2011)}. 
Here we use the largest of these.
This does not affect the analysis presented herein as we are interested here in normalization between disciplines, not between 
subdisciplines.}.
These fits are also depicted in Fig.2(a). 
The disparity between the computer sciences and the other two disciplines is evident.
The plot demonstrates that the  quality measures for the computer sciences plateau at $s \approx \tilde{s}_> = 57.5$
while those for the biology and physics  peak at  $s \approx \tilde{s}_> = 37.1$ and $40.2$, respectively. 

Thus the computer science teams appear to have performed significantly better at RAE than both physics and biology,
which are comparable with each other.
On the other hand, according to the analysis of \red{King (2004)}, where the strength of research was compared 
across nations, the UK is particularly strong in biology when compared internationally.
If the RAE results as displayed in Fig.2(a) are taken literally, the biggest and best computer science
teams in the UK are performing at levels about $50\%$ above the biggest and best physics and biology teams,  
which themselves are amongst the strongest internationally. 
A more likely explanation for the misalignment apparent in Fig.2(a) is a higher degree of stringency in the RAE
evaluation panels for physics and biology than for computer science.
This conclusion is reinforced in Fig.2(b), where the fits (\ref{Nc}) are compared for a variety of different 
research areas. 
To facilitate comparison between them, the areas featured here are those with relatively large 
critical masses $N_c$ and small slopes $b_2$ on the right.
Clearly then, these results need to be normalized to facilitate fair and meaningful comparison between
teams across disciplines. 

For these disciplines, the plateaus in the fits in the supercritical regions reflect the best research qualities achievable,
on average, by large research teams. 
As stated above, an obvious and sensible way to normalize quality scores, then,  is to adjust them in such a way 
that the plateaus in Fig.2(b) are at similar levels. 
To address the question of how to achieve this, we plot a number of statistics in the right part
of Fig.2(b) to determine which, if any, captures the characteristics of the plateaus for the different disciplines. 
The notation is as follows:
\begin{eqnarray*}
 \bar{s} & = & {\mbox{mean quality of all groups in given discipline}} \\
 \tilde{s} & = & {\mbox{mean value weighted by the number of individuals in each group}} \\
 a_2     & = & {\mbox{intercept of right fitted line through the $s$ axis}} \\
 s_c     & = & {\mbox{fitted quality value at $N=N_c$ for the given discipline}}\\
 \bar{s}_>    & = & {\mbox{mean quality value for large groups ($N>N_c$)}} \\
 \tilde{s}_>  & = & {\mbox{mean quality value, weighted by group size, for large groups}}  
\end{eqnarray*}
A sensible normalization scheme should adjust the plateaus of Fig.2 so that they are of similar level.
The last two statistics, $\bar{s}_>$ and $\tilde{s}_>$ appear to best capture both the spread in the heights of
the plateaus and the relative heights of each plateau for each discipline\footnote{
For example, the inappropriateness of the overall means $\bar{s}$ or $\tilde{s}$ is clear from Fig.2(b). They would each 
rank the education UOA as lowest and the medical sciences UOA as highest within the set even though the fits to the 
supercritical groups for both of these disciplines lie centrally within the mass of such fits to other UOA's in Fig.2(b).
}.

\begin{figure}[t]
\begin{center}
\includegraphics[width=0.49\columnwidth, angle=0]{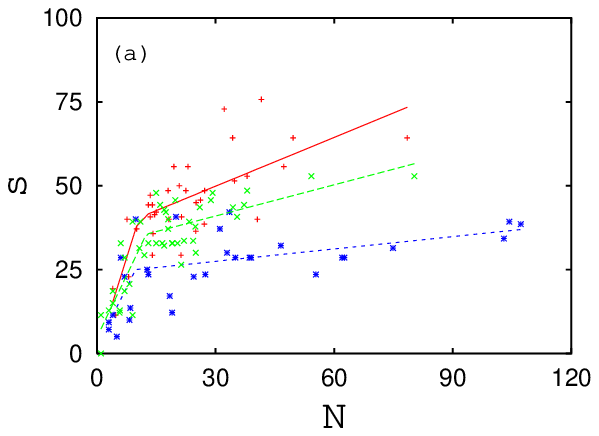}
\includegraphics[width=0.49\columnwidth, angle=0]{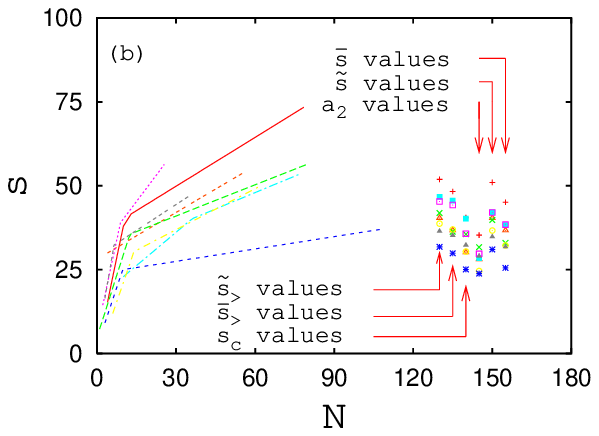}
\caption{(a) Plots of team quality versus quantity for economics \& econometrics~({\color{red}{+}}),
applied mathematics~({\color{green}{$\times$}}), 
and agriculture, veterinary and food sciences~({\color{blue}{\textasteriskcentered}}),
together with the best piecewise linear fits. 
(b) The corresponding fits for
history of arts, performing arts, communication, \& media studies~({\color{magenta}{$\Box$}}),
French, German, Dutch and Scandinavian languages~({\color{Gray}{$\blacktriangle$}}),
pure mathematics~({\color{BurntOrange}{$\triangle$}}),
chemistry~({\color{cyan}{$\blacksquare$}}),
sociology~({\color{yellow}{O}}),
and various statistics for these disciplines.
}
\end{center}
\end{figure}

Therefore the supercritical weighted means $\tilde{s}_>$  offer a sensible measure upon which to base normalization across disciplines
and these are listed in Table~1. 
The average of the  values listed  is $\hat{s}_> = 43.2$. 
The data for each discipline are now reweighted as 
\begin{equation}
 s \rightarrow s_{\rm{norm}} = \frac{\hat{s}_>}{\tilde{s}_>} s.
\label{norm}
\end{equation}
The discipline-dependent normalization  factors $\hat{s}_>/\tilde{s}_>$ are  listed 
in the final column of Table~1. These are the amounts by which the RAE quality scores have to be adjusted to correct for the 
differences in stringency in peer evaluations across disciplines. These factors therefore provide the answer to the call 
by Higgins for an ``intellectual basis'' for normalization  discussed  by \red{Corbyn (2009)} and summarized in the Introduction.

The weighted means of the normalized scores $s_{\rm{norm}}$ for supercritical teams in the various 
disciplines are now all coincident at $\hat{s}_> = 43.2$.
The resulting reweighted data together with the corresponding reweighted fits are plotted in Fig.3(a) 
for the three disciplines depicted in Fig.2(a).
In Fig.3(a), the computer science results are now better aligned with those from the other two disciplines,
and there is a greater degree of overlap between the quality scores for the three disciplines. 
Moreover, the reasonable alignment between physics and biology evident in Fig.2(a) is not adversely affected by 
normalization process (in fact it is improved) and the strength of the best biology teams remain strong post normalization.
In Fig.3(b) the normalized fits for the subject  areas of Fig.2(b) are plotted.
Indeed, both plots in Fig.3 appear to offer a fairer representation than their counterparts in Fig.2
and therefore facilitate meaningful comparison across different academic disciplines.

\begin{figure}[t]
\begin{center}
\includegraphics[width=0.49\columnwidth, angle=0]{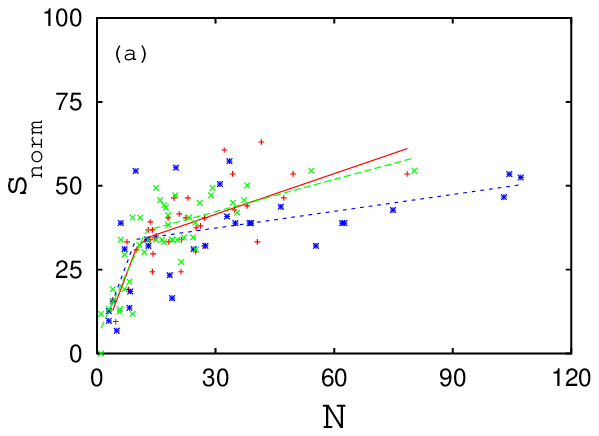}
\includegraphics[width=0.49\columnwidth, angle=0]{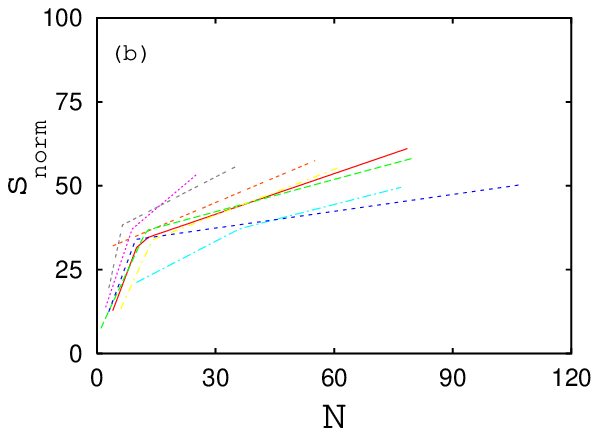}
\caption{(a) The result of the normalization procedure on the three sets of data and corresponding fits  of Fig.4(a).
(b) The post-normalization fits for the disciplines of Fig.4(b).}
\end{center}
\end{figure}

Most of the remaining disciplines listed in Table~1 have relatively small critical masses and therefore, while the
dependency of quality on quantity reduces in entering the supercritical zone, the
slopes of the linear fits to the right of $N_c$ are non-zero ($b_2 >0$) for these subjects.
The raw RAE quality scores $s$ are plotted against team size for three such disciplines
(economics and econometrics, applied mathematics and agriculture, veterinary and food sciences) in Fig.4(a).
The piecewise linear fits coming from model (\ref{Nc}) are also plotted in each case.
Again, there is a clear misalignment between disciplines, with economics appearing to perform far better than 
agriculture and applied mathematics in the middle. 
A plot of the fitted curves for a variety of such disciplines is given in Fig.4(b).
(We have included the anomalous case of chemistry, which has large $N_c$ but non-zero $b_2$, in this plot.
This case is discussed in \red{Kenna and Berche (2010)}.)

Reweighting according to the process (\ref{norm}), one obtains the plots depicted in Fig.5.
In Fig.5(a), the fits for economics and applied mathematics are now better aligned. That for agriculture
remains beneath the former two, but the overlap in data for individual teams appears improved.
Similarly, the fits in Fig.5(b) have a reduced degree of splay when compared with their raw counterparts of
Fig.4(b). 
Again, this has the advantage that the average qualities of large groups in each discipline are 
rendered more similar.

\section{Conclusions}

Assessment systems such as the UK's RAE use expert evaluation to perform comparisons 
between research teams within given academic disciplines. 
However, the absence of a method to compare {\emph{across}} disciplines has been a fundamental flaw
of such exercises, and calls have been issued to remedy this flaw \red{(Corbyn, 2009)}. This paper is a response to such calls.
Normalization across disciplines is required to compensate for different degrees of stringency in the 
expert evaluation within disciplines and is essential for meaningful comparison between research groups in different areas.
Since the British system is heavily reliant on the results of the RAE, an uncomplicated, robust normalization system 
is also essential in order to ensure fair allocation of finances by funding councils.

A simple normalization of quality scores on the basis of overall means (i.e., rescaling the quality measurements for 
each discipline in such a way that the normalized scores have similar means) does not allow
for the fact that different disciplines may have different strengths within a given country.
Also, although techniques exist to normalize bibliometrics and citation counts across disciplines, these are inappropriate
for the RAE and are not used by the British funding agencies. 
This is because bibliometrics directly measure impact rather than quality and
the former are not uniformly good proxies for the latter across all groups and  disciplines.

A more sophisticated but simple method has been proposed here. The notion of critical mass in research is integral to this approach.
For disciplines with relatively large critical masses, the quality of research of large teams tends to plateau and
this phenomenon presents a basis for normalization. 
Since the  average quality of such large teams approximates the best one may expect in a given discipline,
it is sensible to normalize quality measurements across disciplines in such a way that they have 
similar plateaus. 
Application of the same approach to subject areas which have smaller critical masses and which do not exhibit
such plateaus, also reduces the differences in splays of quality scores in different disciplines
and aligns the results in what appears to be an improved manner. 
Thus, such an approach offers a general basis for inter-disciplinary normalization,
while intra-disciplinary discrimination between small, medium and large research teams 
may continue to be performed by subject experts.

\bigskip

\vspace{1cm}
\noindent
{\bf{Acknowledgements}} 
We thank  David Arundel, Christian von Ferber, Neville Hunt, and Housh Mashhoudy for inspiring discussions.

%

\begin{description}

\item[]
Alonso, S, F~J~Cabrerizo, E~Herrera-Viedma and F~Herrera 2009
h-Index: a review focused in its variants, computation and standardization for different scientific fields.
{\emph{Journal of Informetrics}\/},
3,  273-289.

\item[]
Corbyn, Z 2009
End unfit comparisons. 
{\emph{Times Higher Education}\/},
1926,  20.

\item[]
Costas, R, M~Bordons, T~N~van~Leeuwen and A~F~J~van~Raan 2009
Scaling rules in the science system: influence of field-specific citation characteristics on the impact of individual researchers.
{\emph{Journal of the American Society for Information Science and Technology}\/}, 
60 740-753.

\item[]
Evidence (Thompson Reuters) 2009
{\emph{The use of bibliometrics to measure research quality in UK higher education institutions}\/}.
Report to Universities UK \\
$\langle$www.universitiesuk.ac.uk/Publications/Documents/bibliometrics.pdf$\rangle$,
last accessed 8 January 2011.


\item[]
Harnad, S 2008
Validating research performance metrics against peer rankings.
{\emph{Ethics in Science and Environmental Politics}\/},
8,  103-107.

\item[]
Harnad, S 2009
Open access scientometrics and the UK Research Assessment Exercise.
{\emph{Scientometrics}\/},
79, 147-156.

\item[]
Harrison, M 2009
Does high quality research require critical mass?
in 
{\emph{ The Question of R\&D Specialisation: Perspectives and Policy Implications}\/}, 
ed. D~Pontikakis, D~Kriakou and R~van~Baval.
European Commission: JRC Technical and Scientific Reports pp~57-59.

\item[]
Ingham, A~G, G~Levinger, J Graves and V Peckham 1974
The Ringelmann effect: studies of group size and group performance.
{\emph{Journal of Experimental Psychology}\/},
10, 371-384.

\item[]
Katz, J~S 1999
The self-similar science system.
{\emph{Research Policy}\/}, 
28, 501-517.

\item[]
Katz, J~S 2000
Scale-independent indicators and research evaluation.
{\emph{Science and Public Policy}\/}, 
27 23-36.

\item[]
Katz, J~S 2005
Scale-Independent Bibliometric Indicators.
{\emph{Measurement: Interdisciplinary Research and Perspectives}\/}, 
3, 24–28.

\item[]
Kenna,~R and B~Berche 2010
The extensive nature of group quality.
{\emph{Europhysics Letters}\/},  
90 58002.

\item[]
Kenna,~R and B~Berche 2011
Critical mass and the dependency of research quality on group size. 
{\emph{Scientometrics}\/},
86, 
527-540.

\item[]
King, D~A 2004
The scientific impact of nations.
{\emph{Nature}\/},  
430,  311-316.

\item[]
Leydesdorff~L and L~Bornmann 2010
How fractional counting of citations affects the Impact Factor: Normalization in terms of differences in citation potentials among fields of science.
{\emph{Journal of the American Society for Information Science and Technology}\/},
DOI: 10.1002/asi.21450.

\item[]
Leydesdorff,~L and T~Opthof 2010
Normalization at the field level: Fractional counting of citations.
{\emph{Journal of Informetrics}\/}, 
4 644-646.

\item[]
Leydesdorff~L and J~C~Shin 2010
How to evaluate universities in terms of their relative citation impacts:
      Fractional counting of citations and the normalization of differences among disciplines.
{\emph{arXiv:1010.2465}\/}.

\item[]
Mahdi, S, P~D'Este and A~Neely 2008
Citation counts: are they good predictors of RAE scores? A Bibliometric Analysis of RAE 2001.
$\langle$http://ssrn.com/abstract=1154053$\rangle$, last accessed 8 January 2011.

\item[]
Merton, R~K 1968
The Matthew Effect in Science. 
{\emph{Science}\/}, 
159,  56-63.

\item[]
Oppenheim~C and M~A~C~Summers 2008
Citation counts and the Research Assessment Exercise part VI: Unit of Assessment 67 (music).
{\emph{Information Research}\/}, 
13, 2.

\item[]
Opthof, T and L~Leydesdorff 2010
Caveats for the journal and field normalizations in the CWTS (``Leiden'') evaluations of research performance.
{\emph{Journal of Informetrics}\/}, 
4 423-430.

\item[]
Ringelmann, M 1913
Recherches sur les moteurs anim{\'{e}}s: travails de l'homme. 
{\emph{Annales de l'Institut National Agronomique}\/}, 
2, 2-39.

\item[]
Tsay, M-Y 2009
An analysis and comparison of scientometric data between journals of physics, chemistry and engineering.
{\emph{Scientometrics}\/},
78, 279-293.

\item[]
van~Raan, A~F~J 2005
Fatal attraction: Conceptual and methodological problems in the ranking of universities by bibliometric methods.
{\emph{Scientometrics}\/},
62, 133-143.

\item[]
van~Raan,~A~F~J 2006a
Comparison of the Hirsch-index with standard bibliometric indicators and with peer judgement for 147 chemistry research groups.
{\emph{Scientometrics}\/}, 
67,  491-502.

\item[]
van Raan, A~F~J 2006b 
Statistical properties of bibliometric indicators: research group indicator distributions and correlations.
{\emph{Journal of the American Society for Information Science and Technology}\/},
57  408-430.

\item[]
van Raan, A~F~J 2006c
Performance-related differences of bibliometric statistical properties of research groups: cumulative advantages and hierarchically layered networks.
{\emph{Journal of the American Society for Information Science and Technology}\/},
57, 1919-1935.

\item[]
van~Raan, A~F~J 2006d
Statistical Properties of Bibliometric Indicators: Research Group Indicator Distributions and Correlations.
{\emph{Journal of the American society of Information Science and Technology}\/},
57, 408–430.

\item[]
van Raan, A~F~J 2008
Bibliometric statistical properties of the 100 largest European research universities: prevalent scaling rules in the science system.
{\emph{Journal of the American Society for Information Science and Technology}\/}, 
59, 461-475.

\item[]
van~Raan, A~F~J, T~N~van Leeuwen, M~S~Visser, N~J~van~Eck and L~Waltman 2010
Rivals for the crown: Reply to Opthof and Leydesdorff.
{\emph{Journal of Informetrics}\/},
4 431-435.

\item[]
Waltman, L, N~J~van~Eck, T~N~van~Leeuwen, M~S~Visser and A~J~van~Raan 2010
Towards a new crown indicator: Some theoretical considerations.
{\emph{Journal of Informetrics}\/}, 
DOI: 10.1016/j.joi.2010.08.001.

\end{description}

\end{document}